\documentclass[aps,prl,preprint,groupedaddress,amsmath,amssymb,amsfonts,nofootinbib]{revtex4-1}

\usepackage{amssymb,amsfonts}
\usepackage[all,arc]{xy}
\usepackage{enumerate}
\usepackage{mathrsfs}
\usepackage{graphicx}

\begin{document}

\bibliographystyle{plain}

\title{Retarded Fields of Null Particles and the Memory Effect}

\author{Alexander Tolish}
\email{tolish@uchicago.edu}
\author{Robert M. Wald}
\email{rmwa@uchicago.edu}
\affiliation{Enrico Fermi Institute and Department of Physics \\
  The University of Chicago \\
  5640 S. Ellis Ave., Chicago, IL 60637, U.S.A.}

\begin{abstract}
We consider the retarded solution to the scalar, electromagnetic, and linearized gravitational field equations in Minkowski spacetime, with source given by a particle moving on a null geodesic. In the scalar case and in the Lorenz gauge in the electromagnetic and gravitational cases, the retarded integral over the infinite past of the source does not converge as a distribution, so we cut off the null source suitably at a finite time $t_0$ and then consider two different limits: (i) the limit as the observation point goes to null infinity at fixed $t_0$, from which the ``$1/r$'' part of the fields can be extracted and (ii) the limit $t_0 \to - \infty$ at fixed ``observation point.'' The limit (i) gives rise to a ``velocity kick'' on distant test particles in the scalar and electromagnetic cases, and it gives rise to a ``memory effect'' (i.e., a permanent change in relative separation of two test particles) in the linearized gravitational case, in agreement with previous analyses. 
As already noted, the second limit
does not exist in the scalar case or for the Lorenz gauge vector potential and Lorenz gauge metric perturbation in the electromagnetic and linearized gravitational cases. However, in the electromagnetic case, we obtain a well defined distributional limit for the electromagnetic field strength, and in the linearized gravitational case, we obtain a well defined distributional limit for the linearized Riemann tensor. In the gravitational case, this limit agrees with the Aichelberg-Sexl solution. There is no ``memory effect'' associated with this limiting solution. 
This strongly suggests that the memory effect---including nonlinear memory effect of Christodoulou---should not be interpreted as arising simply from the passage of (effective) null stress energy to null infinity but rather as arising from a ``burst of radiation'' associated with the creation of the null stress-energy (as in case (i) above) or, more generally, with radiation present in the spacetime that was not ``produced'' by the null stress-energy.

\end{abstract}
\maketitle

\section{Introduction}
As is well known, gravitational radiation induces relative displacements in a system of inertial test particles.
Zel'dovich and Polnarev \cite{Zeldovich} first noted that, within linearized gravity, the passage of a pulse of gravitational radiation can cause a permanent change in the relative displacement these particles. This effect is known as the {\it memory effect}. The net relative displacement, $\Delta D^a$, after passage of the pulse for test masses with initial separation $D^a$ can be expressed as 
\begin{equation}
\Delta D_a = \frac{1}{2}\Delta h^{\textnormal{TT}}_{ab} D^b \, ,
\end{equation}
where $\Delta h^{\textnormal{TT}}_{ab}$ is the net change in the metric perturbation in a transverse-traceless gauge. In linearized gravity, for gravitational radiation produced by a change in the motion of particle-like sources occurring in a localized region of spacetime, to leading order in $1/r$ we have \cite{Thorne}, \cite{WisemanWill} 
\begin{equation}
\Delta h^{\textnormal{TT}}_{ab}= \frac{1}{r} \Delta\sum_A \frac{4M_A}{\sqrt{1-v_A^2}}\left[\frac{(v_A)_a (v_A)_b}{1-v_A\cos\theta_A}\right]^{\textnormal{TT}} \, ,
\label{nonrev}
\end{equation}
where $A$ is an index labeling the source particles, which have mass $M_A$ and velocity $v_A$ at angle $\theta_A$ with respect to the direction of the detector. The brackets $[...]^{\textnormal{TT}}$ represent the transverse-traceless part of the object within, and $\Delta$ means the difference between the quantity at late and early times. In the case where there is an emission of a null-particle-like source of energy $E$ moving in the $z$-direction and all other sources (before and after the emission) are non-relativistic, the memory effect for a detector at $(\theta, \phi)$ becomes
\begin{equation}
\Delta D_a=\frac{E}{r}\frac{\sin^2\theta}{1-\cos\theta}\left(\theta_a \theta_b-\phi_a \phi_b \right)D^b \, ,
\label{nullmem}
\end{equation}
where $\theta^a$ and $\phi^a$ denote unit vectors in the $\theta$ and $\phi$ directions, respectively.

Making use of a careful analysis of the nonlinear Einstein equation, Christodoulou \cite{Christodoulou} found that there can be significant nonlinear contributions to the memory effect. Christodoulou's formula for the nonlinear contribution to the memory effect associated with
the passage of a gravitational radiation to future null infinity is expressed in terms of an integral of the Bondi flux over future null infinity. For the case where the Bondi flux is highly localized 
in the $z$-direction and the integrated flux is $E$, Christodoulou's formula reduces to \eqref{nullmem}.

Thorne \cite{Thorne} and Wiseman and Will \cite{WisemanWill} soon interpreted Christodoulou's nonlinear memory effect as simply corresponding to the linear memory effect, but with the nonlinear effective stress-energy of gravitational waves replacing the ``particle'' sources. Further support for this interpretation can be found in the fact that a similar nonlinear memory effect occurs when a flux of electromagnetic radiation reaches null infinity \cite{Bieri}, thereby showing that the nonlinear memory effect is not special to gravitational waves. Very recently, Bieri and Garfinkle \cite{BG1} have shown that the linear memory effect for null matter can be derived in close parallel to Christodoulou's derivation, thus further confirming that the nonlinear memory effect can be interpreted as being the same as the linear memory effect, with the effective stress energy of gravitational radiation replacing the ordinary stress-energy of null matter.

However, there remains a puzzling aspect of the alternative derivations of the memory effect. In the derivation of the formula \eqref{nonrev} for the linear memory effect, one considers the retarded solution associated with sources. The linear memory effect is thereby seen to be simply an aspect of the gravitational radiation emitted by the sources. In particular, for slowly moving sources, what is relevant for producing a nontrivial memory effect is the net {\em change} in the time derivative of the quadrupole moment of the sources. If the time derivative of the quadrupole moment does not vary---such as for the case of a single particle moving on a timelike geodesic of Minkowski spacetime---there is no gravitational radiation and no memory effect. Similar results hold if one does not assume slow motion of the sources \cite{WisemanWill}.

By contrast, in Christodoulou's \cite{Christodoulou} derivation of the nonlinear memory effect and in the Bieri and Garfinkle \cite{BG1} derivation of the linear memory effect for null matter, there is no allusion to emission by sources. If one examines these derivations, it would appear that all that is relevant to the memory effect is that there be a flux of gravitational radiation or null matter to future null infinity. This flux could just as well have originated from past null infinity as have been emitted by sources at some finite time. This would suggest\footnote{Neither Christodoulou \cite{Christodoulou} nor Bieri and Garfinkle \cite{BG1} propose an interpretation of the ``cause'' of the memory effect.} that the null memory effect should be interpreted as being associated with simply the passage of null (effective) stress-energy to infinity---perhaps as a ``tidal effect''---rather than as being caused by a burst of radiation associated with some ``emission event'' within the spacetime.

We will attempt to gain insight into this issue by considering the simple problem of obtaining retarded solution to the wave equation in Minkowski spacetime with source given by a particle moving on a null geodesic. This directly yields the retarded solution of a scalar field with a null particle scalar charge source, and it also yields the corresponding retarded solution for the electromagnetic and linearized gravitational cases for appropriate choices of gauge. We now summarize our main results, after which we explain the relevance of our results to the interpretation of the memory effect.

First, we find in section 2 that the retarded solution to the wave equation in Minkowski spacetime with a delta-function source on a complete null geodesic does not exist as a distribution. This difficulty is not due to the ``nullness" of the source but rather to its non-compactness. However, if we ``cut off" the source at a time $t_0$ in the past, i.e., if we consider the scalar charge source
\begin{align}
S_0=q\delta(x)\delta(y)\delta(z-t)\Theta(t-t_0) \, ,
\label{scalsource}
\end{align}
the retarded solution is
\begin{equation}
\varphi_0=\frac{q}{u}\Theta\left((t-t_0)-\sqrt{x^2+y^2+(z-t_0)^2}\right) \, ,
\label{phiret}
\end{equation}
where
\begin{equation}
u \equiv t -z
\end{equation}
The right side of \eqref{phiret} is well defined as a distribution. This scalar field associated with a null source created at time $t_0$ will produce a force on a test particle of scalar charge $Q$ given by $f^a=Q\partial^a\varphi_0$. Differentiation of the $\Theta$-function in \eqref{phiret} will yield a
$\delta$-function term in $f^a$ that will give the test particle a $4$-momentum ``kick,'' which can be understood as being due to the radiation produced by the creation of the source at time $t_0$.

Although the limit as $t_0 \to -\infty$ of expression \eqref{phiret} does not exist as a distribution, we shall show in section 2 that $\lim_{t_0 \to -\infty} k^{[a} \partial^{b]} \varphi$
is well defined as a distribution, where $k^a = t^a + z^a$ is the vector field that is parallel to the tangent to the source worldline. Thus, although the retarded field produced by a scalar charge moving on a (past and future complete) null geodesic is ill defined, we obtain a well defined $4$-momentum kick, modulo multiples of $k^a$, from such a source. However, this kick differs from the kick resulting from a creation event at a finite time $t_0$.

As we shall see in section 3, the situation in the electromagnetic case is similar. Maxwell's equations in Lorenz gauge reduces to four scalar wave equations, so we can immediately write down the retarded solution for the vector potential $A_a$ in terms of the retarded solution to the scalar wave equation. However, one important difference is that Maxwell's equations require conservation of charge, so one cannot simply create a charge at a finite time $t_0$, as in \eqref{scalsource}. Nevertheless, one can consider a charge that sits ``at rest'' until time $t_0$ and thereafter moves on a null geodesic. To order $1/r$, the radiation from this sharp change in the $4$-velocity of the source produces an electromagnetic field of the form
\begin{equation}
F^{ab}=-2q\frac{1}{r}\frac{\sin\theta}{1-\cos\theta}\theta^{[a}K^{b]}\delta(U) \, ,
\label{EMcutoff}
\end{equation}
where $K^a=t^a+r^a$ and $U$ is given by eq.~\eqref{upsilon} below. This field, in turn, will produce a ``velocity kick'' on a distant test particle, in agreement with recent results of Bieri and Garfinkle \cite{BieriGarfinkle}.

In section 3, we also consider the limit as $t_0 \to -\infty$ in the electromagnetic case. The limit of the vector of the vector potential in Lorenz gauge does not exist, but the limit of the field tensor $F^{ab} = 2 \nabla^{[a} A^{b]}$ does exist, and we find
\begin{equation}
F^{ab}=-4q\frac{1}{\rho}\rho^{[a}k^{b]}\delta(u) \, ,
\label{EMnocutoff}
\end{equation}
where $\rho = (x^2 + y^2)^{1/2}$ and $\rho^a = \nabla^a \rho$.
This agrees with results of Jackiw, Kabat and Oritz \cite{JKO}. This retarded field of a charged particle that moves on a null geodesic forever also gives rise to a velocity kick on a distant test particle, but this kick is very different from the velocity kick produced by the field \eqref{EMcutoff}. 

In section 4, we treat the linearized gravitational case. Again, we can immediately write down the solution to the linearized Einstein equation in Lorenz gauge in terms of the retarded solution to the scalar wave equation. However, the linearized Einstein equation requires conservation of $4$-momentum of the source, so we can neither create a mass at time $t_0$ nor have a mass initially ``at rest'' suddenly start moving on a null geodesic. Nevertheless, we can start with a particle of mass $M$ ``at rest'' and, at time $t_0$, have it ``emit'' a particle of energy $E$ that moves on a null geodesic, with the original particle then losing mass and recoiling so as to conserve total $4$-momentum. As we shall see in section 4, to order $1/r$ and to leading order in $E/M$, the Riemann tensor of the retarded solution is
\begin{align}
R_{abcd}=4E&\frac{1}{r}\big[\frac{2}{1-\cos\theta}k_{[a}K_{b]}K_{[c}k_{d]}\nonumber\\
&-\left(2K_{[a}(t_{b]}z_{[c}+z_{b]}t_{[c})K_{d]}+(1+\cos\theta)\left(2K_{[a}t_{b]}t_{[c}K_{d]}+K_{[a}\eta_{b][c}K_{d]}\right)\right)\big]\delta'(U) \, .
\label{Cutoff}
\end{align}
This Riemann tensor produces a ``relative displacement kick'' on test particles of the form
\begin{equation}
\Delta D_a=E\frac{1}{r}\frac{\sin^2\theta}{1-\cos\theta}\left(\theta_a \theta_b-\phi_a\phi_b\right)D^b \, ,
\end{equation}
in agreement with the form of the memory effect for null matter.

In section 4, we also take the limit as $t_0 \rightarrow - \infty$. Although the metric perturbation in Lorenz gauge does not exist as a distribution in this limit, the linearized Riemann tensor has the limit
\begin{equation}
R_{abcd}=16E\frac{1}{\rho^2}k_{[a}(\rho_{b]}\rho_{[c}-\phi_{b]}\phi_{[c})k_{d]}\delta(u) \, - \, 16\pi Ek_{[a}q_{b][c}k_{d]}\delta(x)\delta(y) \delta(u) \, ,
\label{NoCutoff}
\end{equation}
where $q_{ab}$ is the projection of the metric into the ``$x-y$'' plane. Thus, the Riemann tensor \eqref{NoCutoff} corresponds to the retarded linearized curvature produced by a null particle. Eq. \eqref{NoCutoff} agrees with the Aichelberg-Sexl solution \cite{AichelburgSexl} (modulo what appear to be some sign misprints in their eq.(3.12)). The Riemann tensor \eqref{NoCutoff} has no ``derivative of a $\delta$-function'' piece, so unlike \eqref{Cutoff}, it provides no ``relative displacement kick'' to test particles. However, it does provide a ``relative velocity kick'' to test particles, which falls off as $1/r^2$. Interestingly, as we shall show in section 4, this instantaneous relative velocity kick agrees, up to a factor of $2$, with the integrated relative velocity change of test particles that would occur in Newtonian gravity due to tidal effects produced by the passage of a particle of mass $m=E$ moving with velocity $v=c$. Thus, the Aichelberg-Sexl Riemann tensor may be thought of as corresponding to a ``special relativistic compression'' of the Newtonian tidal effects of a particle moving at the speed of light into the null hyperplane containing the null particle source.

Returning, finally, to the questions that motivated our investigations, we see that the Riemann tensor \eqref{NoCutoff} represents the retarded field produced by a null particle source in linearized gravity. As just noted above, this is a ``pure tidal field'' and there is no memory effect associated with this tidal field. We conclude that the memory effect should not be interpreted as being ``caused by'' the passage of null (effective) stress-energy to infinity. Conversely, the fact that there is 
a memory effect associated with the passage of null (effective) stress-energy to infinity is directly related to the fact that the Riemann tensor \eqref{NoCutoff} is not physically acceptable: It fails to be asymptotically flat at spatial infinity (even if we ``smooth out'' the source, as we can in linearized gravity), since the Riemann tensor vanishes in all non-equatorial directions and falls off too slowly (as $1/r^2$) in equatorial directions near spatial infinity. In order to have a solution with a null source that is asymptotically flat at spatial infinity, one must either ``emit'' the null source at a finite time or have ``additional radiation'' incoming from infinity. The memory effect should be thought of as being produced by the gravitational radiation resulting from such an emission event or such additional radiation.

\section{Scalar Field}

As discussed in the previous section, we are interested in obtaining the retarded solution to the massless scalar wave equation in Minkowski spacetime 
\begin{equation}
\nabla^a \nabla_a \varphi = -4\pi S
\label{waveeq}
\end{equation}
where the source, $S$, corresponds to a (scalar) charged particle moving on a null geodesic, which
we take to be moving in the ``$z$-direction'' in some global inertial coordinates $(t,x,y,z)$
\begin{equation}
S(t,x,y,z) = q \delta(x)\delta(y)\delta(z-t) \, .
\label{source}
\end{equation}
We will denote events in spacetime by capital letters (i.e., $X$ and $X'$), so as not to confuse spacetime points with the $x$-coordinate of our global inertial coordinates.

We note, first, that although there appears to be a widespread belief that charged particle sources that move at the speed of light (or faster than light) should somehow be ``illegal'' (see, e.g., the remark below eq.(2.7) of \cite{AichelburgSexl}), there is, in fact, no difficulty in obtaining the retarded solution (as a distribution) to the wave equation \eqref{waveeq} for any distributional source, $S$, of compact support. This can be seen as follows: If $S$ is of compact support, the problem of obtaining the (distributional) retarded solution, $\varphi_R(X)$, to \eqref{waveeq} is essentially the same as defining the product of the distributions $G_R(X,X')$ and $S(X')$, where $G_R$ denotes the retarded Green's function,
\begin{equation}
G_R(t,\vec{x};t',\vec{x}')=\frac{1}{2\pi}\delta \left[-(t-t')^2+|\vec{x}-\vec{x}'|^2 \right] \Theta(t-t') \, ,
\end{equation}
since for any test function, $f$, we have $\varphi_R(f) = G_R S(F)$,
where $F(X,X') = f(X) h(X')$, with $h$ being any test function with $h=1$ on the support of $S$. As a distribution on ${\bf R}^4 \times {\bf R}^4$, the wavefront set of $G_R(X,X')$ is known to be of the form \cite{dh}, \cite{rad}
\begin{equation}
{\rm WF} [G_R] = \{(X,K;X',-K')\}
\end{equation}
where $X$ lies on a future-directed null geodesic starting from $X'$, $K$ is a (future- or past-directed) (co-)tangent to this geodesic\footnote{For $X=X'$, the wavefront set is $\{X,K;X,-K\}$ for all $K \neq 0$.} at $X$, and $K'$ is the parallel transport of $K$ to $X'$. As a distribution on ${\bf R}^4 \times {\bf R}^4$, the wavefront set of $S(X')$ is of the form 
$\{(X,0;X',K')\}$ where $(X',K')$ is in the wavefront set of $S$ as a distribution on ${\bf R}^4$. Hence, we cannot get a zero cotangent vector in ${\bf R}^4 \times {\bf R}^4$ by adding cotangent vectors in ${\rm WF} [G_R]$ to those in ${\rm WF} [S]$. It follows (see, e.g., \cite{rs}) that $G_R S$ is well defined as a distribution for any distribution $S$, and the retarded solution is well defined as a distribution for any $S$ of compact support. Note that this argument generalizes straightforwardly to an arbitrary globally hyperbolic curved spacetime.

On account of the support properties of $G_R$, it is obvious that the requirement that $S$ be of compact support can be replaced by the requirement that $S$ vanish to the past of some Cauchy surface. However, the source \eqref{source} does not have this property, so it is not obvious that the retarded solution exists. Consequently, we will, instead, consider the source
\begin{equation}
S_0 (t,x,y,z) = q \delta(x)\delta(y)\delta(z-t)\Theta(t-t_0) \, .
\label{source2}
\end{equation}
corresponding to the ``creation'' of a scalar charged particle at time $t_0$, which subsequently moves on a null geodesic. By the above general arguments, the retarded solution with source \eqref{source2} exists as a distribution. We will then consider the limit $t_0 \to -\infty$.

The retarded solution with source \eqref{source2} is
\begin{eqnarray}
\varphi_0(X)&=&4\pi\int\textnormal{d}^4x'\;G_R(X; X')S_0(X')\nonumber\\
&=&2q\int\;\textnormal{d}^4x'\;\delta\left[-(t-t')^2+ |\vec{x}-\vec{x}'|^2\right]\Theta(t-t') \delta(x')\delta(y')\delta(z'-t')\Theta(t'-t_0) \, .
\end{eqnarray}
Carrying out the $\delta$-function integrations over $x',y',z'$, we obtain
\begin{eqnarray}
\varphi_0(X)
&=&2q\int\;\textnormal{d}t'\;\delta\left[-(t-t')^2+x^2+y^2+(z-t')^2\right]\Theta(t-t')\Theta(t'-t_0)\nonumber\\
&=&2q\int\;\textnormal{d}t'\;\delta\left[2(t-z)t'-t^2+x^2+y^2+z^2 \right]\Theta(t-t')\Theta(t'-t_0)\nonumber\\
&=&\frac{q}{t-z}\Theta\left(t-\frac{t^2-x^2-y^2-z^2}{2(t-z)}\right)\Theta\left(\frac{t^2-x^2-y^2-z^2}{2(t-z)}-t_0\right) \, .
\end{eqnarray}
The two step functions can be combined into a single step function to produce our final result
\begin{equation}
\varphi_0=\frac{q}{t-z}\Theta\left((t-t_0)-\sqrt{x^2+y^2+(z-t_0)^2}\right) \, .
\label{retsol}
\end{equation}
Note that although $\varphi_0$ is unbounded (since it diverges as $t \downarrow z$ at $x=y=0$) it is locally in $L^1$ and thus is well defined as a distribution.

The ``$4$-force,'' $f^a$, exerted by the field $\varphi_0$ on a test particle of charge $Q$ is
\begin{align}
f^a=Q\nabla^a\varphi_0 \, .
\end{align}
Unlike the case of electromagnetism, $f^a$ is not automatically orthogonal to the $4$-velocity of the test particle, and hence will, in general, produce a change in the rest mass of the particle as well as a change in its momentum. From \eqref{retsol}, we obtain
\begin{equation}
f^a =\frac{qQ}{u^2}k^a\Theta(U)-\frac{qQ}{u}\left(t^a+\frac{rr^a-t_0z^a}{\sqrt{x^2+y^2+(z-t_0)^2}}\right)\delta(U) \, , 
\label{scalarforce}
\end{equation}
where
\begin{equation}
u = t-z
\end{equation}
\begin{equation}
U=(t-t_0)-\sqrt{x^2+y^2+(z-t_0)^2} \, ,
\label{upsilon}
\end{equation}
and $t^a$ and $r^a$ are unit vectors in the time and radial directions. Note that $U = 0$ corresponds to the future light cone of the event occurring at $t = z= t_0, x=y=0$, where the source \eqref{source2} was created. 
To leading order in $1/r$, our expression \eqref{scalarforce} becomes
\begin{equation}
f^a =- \frac{qQ}{r}\frac{K^a}{1-\cos\theta} \delta(U) + O(1/r^2)\, ,
\end{equation}
where
\begin{equation}
K^a = t^a + r^a \, .
\end{equation}
This $\delta$-function contribution to $f^a$ will give rise to an instantaneous ``kick'' in the $4$-momentum of the test particle. If the test particle is initially ``at rest'' and its motion remains non-relativistic, then the change in $4$-momentum due to this instantaneous kick is given by
\begin{equation}
\Delta P^a =- \frac{qQ}{r}\frac{K^a}{1-\cos\theta}\, . 
\label{momkick}
\end{equation}
Note that this expression for the net kick is independent of $t_0$, i.e., a change in $t_0$ affects the kick only to higher order in $1/r$ 
(although, of course, a change of $t_0$ affects the time at which the kick is felt). Since the kick arises from the $\delta(U)$ term in the force, the kick can be understood as being produced by a burst of radiation emitted when the source was created. Note that the kick diverges as $\theta \to 0$.

Let us now take the limit as $t_0 \to -\infty$. Naively taking the limit of \eqref{retsol}, we obtain
\begin{equation}
\varphi = \lim_{t_0 \to -\infty}\varphi_0=\frac{q}{t-z}\Theta(t-z) \, .
\end{equation}
However, the right side of this equation is not locally in $L^1$ and does not make sense as a distribution. Indeed, it is easy to see that for any fixed, non-negative test function $f$ with $f \neq 0$ at some point at which $t=z$ we have 
\begin{equation}
\lim_{t_0 \to -\infty} \int \varphi_0 f = \infty \, ,
\end{equation}
so the weak distributional limit of $\varphi_0$ does not exist as $t_0 \to -\infty$. We conclude, therefore, that for the scalar wave equation, it does not make sense to talk about the retarded field of a charged particle\footnote{Since, in Minkowski spacetime, averging over the observation point is equivalent to averaging over the source, the failure to obtain a distributional solution for a particle source moving forever on a null geodesic implies the failure to have any retarded solution at all for a smooth, null fluid source with everywhere parallel $4$-velocity.} source that moves forever on a null geodesic. As our derivation has indicated, the problem with obtaining a distributional solution arises from the ``forever" (i.e., non-compactness) character of the source rather than its ``null" character.

Nevertheless, although $\lim_{t_0 \to -\infty} \varphi_0$ does not exist as a distribution, some aspects of this limit do exist. Specifically, let $k^a = t^a + z^a$ be the vector field on Minkowski spacetime that is everywhere parallel to the tangent to the null geodesic source \eqref{source2}. Then we claim that the weak distributional limit $\lim_{t_0 \to -\infty} k^{[a} \nabla^{b]}\varphi_0$ does exist. To see this, let $\alpha^{ab}$ be a smooth, antisymmetric tensor field of compact support. We wish to evaluate 
\begin{equation}
\lim_{t_0 \to -\infty} -\int \varphi_0  k_a \nabla_b \alpha^{ab}\  
= \lim_{t_0 \to -\infty} -\int_{U > 0} {\rm d}^4 x \,\, \frac{1}{u} k_a \nabla_b \alpha^{ab} \, .
\label{smearing1}
\end{equation}
Integrating by parts, we obtain
\begin{equation}
-\int_{U > 0} {\rm d}^4 x \,\, \frac{1}{u} k_a \nabla_b \alpha^{ab} =  -\int_{U > 0} {\rm d}^4 x \,\, \frac{1}{u^2} k_a\nabla_b u \alpha^{ab} - \int_{U = 0}  \frac{1}{u} k_a n_b \alpha^{ab}
\end{equation}
where $n^a$ is the normal to the $U=0$ surface,
\begin{equation}
n^a=t^a+\frac{\rho \rho^a+(z-t_0)z^a}{\sqrt{\rho^2+(z-t_0)^2}} \, .
\end{equation}
(Here $\rho = (x^2 + y^2)^{1/2}$ and $\rho^a = \nabla^a \rho$.)
The bulk integral vanishes because $\alpha^{ab}$ is antisymmetric and $k_{[a}\nabla_{b]}u = 0$. The surface term is
\begin{equation}
- \int_{U = 0}  \frac{1}{u} k_a n_b \alpha^{ab} = -\int_{U = 0} \rho\,\textnormal{d}\rho\,\textnormal{d}\phi\,\textnormal{d}z \frac{\sqrt{\rho^2+(z-t_0)^2}k_{[a}t_{b]}+\rho k_{[a}\rho_{b]}+(z-t_0)k_{[a}z_{b]}}{\left(\sqrt{\rho^2+(z-t_0)^2}-(z-t_0)\right)\sqrt{\rho^2+(z-t_0)^2}} \alpha^{ab}
\end{equation}
As $t_0 \to -\infty$, the numerator in this expression converges uniformly on compact sets to $\rho k_{[a}\rho_{b]}$, whereas the denominator converges uniformly on compact sets to $\rho^2/2$. Furthermore, as $t_0 \to -\infty$, we have $U\to u$. From this it can be seen that the (weak) limit of $k^{[a} \nabla^{b]}\varphi_0$ as $t_0 \to -\infty$ exists and is given by
\begin{equation}
\lim_{t_0\rightarrow-\infty}k^{[a}\nabla^{b]}\varphi_0=2q\frac{1}{\rho}\rho^{[a}k^{b]}\delta(u) \, .
\label{finalscalarforce}
\end{equation}
Thus, in the limit $t_0 \to -\infty$, the force exerted on a test particle is well defined modulo addition of multiples of $k^a$. Since this force also has a $\delta$-function character, it gives rise to a $4$-momentum kick of the form
\begin{equation}
\Delta P_\infty^a  = -2qQ\frac{1}{\rho}\rho^a
\end{equation}
modulo multiples of $k^a$. This $4$-momentum kick is very different in form from the kick \eqref{momkick} produced by the burst of radiation arising from a ``creation event.''

\section{Electromagnetic Field}

In this section, we wish to obtain the retarded solution to Maxwell's equations with a charged particle source moving on a null geodesic. As in the case of the scalar wave equation, in order to have a well defined solution, we would like to ``create'' the source at a finite time $t_0$ and then consider the limit $t_0 \to -\infty$. However, unlike the scalar case, we cannot ``create'' a charge at a finite time because Maxwell's equations require conservation of charge. Therefore, we consider, instead, a situation where a charge sits ``at rest'' until time $t=t_0$ and thereafter moves on a null geodesic, i.e., we take the $4$-current to be 
\begin{equation}
j_0^a = q \delta(x)\delta(y)\left[\delta(z-t_0)\Theta(t_0-t)t^a
                                   +\delta(z-t)\Theta(t-t_0)k^a\right] \, ,
\end{equation}
where $k^a = t^a + z^a$ is tangent to the null geodesic $x=y=0, t=z$.

Maxwell's equations for the vector potential, $A^a$, in Lorenz gauge, $\nabla_a A^a = 0$, take the form of a wave equation \eqref{waveeq} for each global inertial component of $A^a$. Therefore, we can immediately write down the retarded solution in Lorenz gauge using the well known Coulomb solution for the source for $t < t_0$ and using \eqref{retsol} for $t \geq t_0$. We obtain
\begin{equation}
A_0^a=\frac{q}{\sqrt{x^2+y^2+(z-t_0)^2}} \Theta(-U) t^a +\frac{q}{t-z} \Theta(U) k^a
\label{A0}
\end{equation}
where $U$ was defined in \eqref{upsilon} above.

The electromagnetic field tensor is given in terms of the vector potential by $F^{ab} = 2 \nabla^{[a} A^{b]}$. From \eqref{A0}, we obtain
\begin{equation}
(F_0)^{ab} =-2q\frac{1}{r}\frac{\sin\theta}{1-\cos\theta}\theta^{[a}K^{b]}\delta(U) + O(1/r^2)\, .
\label{F0}
\end{equation}
The force on a test particle of charge $Q$ and $4$-velocity $u^a$ is $f_a = Q F_{ab} u^b$.
As in the scalar case, the leading order in $1/r$ contribution to $f_a$ is a $\delta$-function term, which will give the particle an instantaneous momentum kick. In the case of electromagnetism, $f^a$ is automatically orthogonal to $u^a$ and, hence, does not change the rest mass of the test particle, i.e., the particle gets only a ``velocity kick.'' For a test particle that is initially ``at rest'' and whose motion remains non-relativistic, the instantaneous kick in $4$-momentum is given by
\begin{equation}
\Delta P^a =qQ\frac{1}{r}\frac{\sin\theta}{1-\cos\theta}\theta^a \, .
\label{momkickEM}
\end{equation}
This agrees with the velocity kick obtained by Bieri and Garfinkle \cite{BieriGarfinkle}.

Let us now take the limit $t_0 \to -\infty$. The contribution of the first (Coulomb) term in \eqref{A0} clearly goes to zero in this limit. However, apart from the factor of $k^a$, the contribution of the second term in \eqref{A0} is identical to the scalar case, and hence it does {\em not} have a distributional limit. We conclude that the retarded solution for the vector potential of a charged particle that moves forever on a null geodesic does not exist in Lorenz gauge. Nevertheless, since the Coulomb contribution vanishes in the limit, we see that that
\begin{equation}
F_{ab} \equiv \lim_{t_0 \to -\infty}(F_0)_{ab} = -2 \lim_{t_0 \to -\infty} k_{[a} \nabla_{b]} \varphi_0
\end{equation}
with $\varphi_0$ given by \eqref{retsol}. As we showed in the previous section, the limit on the right side of this equation {\em does} exist as a distribution, and we obtain
\begin{equation}
F_{ab} = -4q\frac{1}{\rho}\rho_{[a}k_{b]}\delta(u) \, .
\label{Fret}
\end{equation}
Equation \eqref{Fret} may thus be interpreted as providing the retarded field\footnote{Note that although we showed above that the retarded vector potential in Lorenz gauge does not exist for this solution, one can find other gauges in which a distributional vector potential for the field (\ref{Fret}) can be found; see \cite{JKO} and \cite{Jackson}.} of a charged particle that moves on a null geodesic for all time, in agreement with Jackiw, Kabat and Oritz \cite{JKO} (see also problem 11.18 of the third edition of Jackson \cite{Jackson}).

The field \eqref{Fret} produces an instantaneous momentum kick on a test particle of charge $Q$ (assumed to be initially at rest) given by
\begin{equation}
\Delta P_\infty^a  = 2qQ\frac{1}{\rho}\rho^a \, .
\end{equation}
Again, this differs in form from the momentum kick \eqref{momkickEM} produced by the burst of radiation associated with the instantaneous change of motion of the source at time $t_0$.

\section{Linearized Gravitational Field}

We now turn to the case of linearized gravity, with a source $T_{ab}$ corresponding to a particle moving on a null geodesic. As in the scalar and electromagnetic cases, we would like to ``create'' this particle at time $t_0$ and then take the limit $t_0 \to -\infty$. However, the linearized Einstein equation requires conservation of stress-energy, which, for particle sources, requires conservation of $4$-momentum. Thus, the simplest case to consider would be a particle of mass $M$ which is at rest until time $t_0$, at which time it emits a null particle of energy, $E$, and then loses mass and recoils so as to conserve $4$-momentum. Thus, we consider a stress-energy source of the form
\begin{eqnarray}
T_0^{ab} &=& \delta(x)\delta(y)\big[M \delta(z-t_0)\Theta(t_0-t)t^at^b + \nonumber \\
&&+ M' \delta(z' - t_0) \Theta(t-t_0) t'^a t'^b
                                   + E \delta(z-t)\Theta(t-t_0)k^a k^b \big] \, ,
                                   \label{gravsource}
\end{eqnarray}
where $M'$ and $t'^a$ are chosen so as to conserve $4$-momentum and $z'$ is the global inertial ``$z$-coordinate'' in the frame in which the recoiling particle is at rest. 

We denote the metric perturbation by $h_{ab}$. As is well known, in Lorenz (harmonic) gauge, the linearized Einstein equation for $\bar{h}_{ab} \equiv h_{ab} - \frac{1}{2} h \eta_{ab}$ (where $\eta_{ab}$ is the Minkowski metric and $h = \eta^{ab} h_{ab}$) takes the form 
\begin{equation}
\nabla^c \nabla_c\bar{h}_{ab}=-16\pi T_{ab} \, ,
\end{equation}
yielding a wave equation for each of its global inertial components. We can therefore immediately obtain the retarded solution for $\bar{h}_{ab}$ with source \eqref{gravsource}---and, hence, obtain the retarded solution for $h_{ab}$---as a sum of 3 pieces: (I) a linearized Schwarzschild piece arising from the first (``particle at rest'') term in the source,
\begin{equation}
(h^{\rm I}_0)_{ab}=\frac{2M}{\sqrt{x^2+y^2+(z-t_0)^2}}\left(\eta_{ab}+2t_at_b\right)\Theta(-U) \, ,
\label{hI}
\end{equation}
(II) a boosted Schwarzschild piece arising from the second (``recoiling particle'') term in the source
\begin{equation}
(h^{\rm II}_0)_{ab}=\frac{2M'}{\sqrt{x^2+y^2+(z'-t_0)^2}}\left(\eta_{ab}+2t'_at'_b\right)\Theta(U) \, ,
\label{hII}
\end{equation}
and (III) a piece arising from the third (``null particle'') term in the source
\begin{equation}
(h^{\rm III}_0)_{ab}=\frac{4E}{t-z}k_ak_b\Theta(U) \, .
\label{hIII}
\end{equation}

The linearized Riemann tensor, $R_{abcd}$, associated with metric perturbation $h_{ab}$ is
\begin{equation}
R_{abcd} = 2\nabla_{[a} \nabla_{|[d} h_{c]|b]} \, .
\end{equation}
The leading order in $1/r$ contribution to the linearized Riemann tensor will arise from differentiation of the $\Theta$-functions
appearing in \eqref{hI}-\eqref{hIII}. Assuming $E \ll M$ and keeping only the leading order term in $E/M$, we find
\begin{align}
R_{abcd}=4E&\frac{1}{r}\big[\frac{2}{1-\cos\theta}k_{[a}K_{b]}K_{[c}k_{d]}\nonumber\\
&-\left(2K_{[a}(t_{b]}z_{[c}+z_{b]}t_{[c})K_{d]}+(1+\cos\theta)\left(2K_{[a}t_{b]}t_{[c}K_{d]}+K_{[a}\eta_{b][c}K_{d]}\right)\right)\big]\delta'(U) \, ,
\label{RI+II+III}
\end{align}
where the first term in the square brackets arises from $(h^{\rm III}_0)_{ab}$ and the remaining terms 
arise from $(h^{\rm II}_0)_{ab}$.
Although we have chosen a particular decay/recoil process in order to do these calculations, the details of the process are irrelevant at $O(E/M)$ provided that all of the particles apart from the null particle are non-relativistic, i.e., the details of the decay process would affect \eqref{RI+II+III} only at higher orders in $E/M$.

The linearized Riemann tensor will produce a relative acceleration (i.e., geodesic deviation) for nearby freely falling test particles. If the particles are initially ``at rest'' (i.e., $4$-velocity parallel to $t^a$) and separated by spatial displacement $D^a$, then
\begin{equation}
t^e \nabla_e t^f \nabla_f D^a = -{R_{bcd}}^a t^b t^d D^c
\end{equation}
i.e., in terms of components
\begin{equation}
\frac{d^2 D^\mu}{dt^2}= -{R_{t \nu t}}^\mu D^\nu \, .
\label{relacc2}
\end{equation}
The ``derivative of a $\delta$-function'' terms in the linearized Riemann tensor will therefore produce an instantaneous relative {\em displacement kick} to the test particles. {\em This is precisely the memory effect.} For the Riemann tensor \eqref{RI+II+III}, we obtain
\begin{equation}
(\Delta D_0)_a =\frac{E}{r}\frac{\sin^2\theta}{1-\cos\theta}\left(\theta_a\theta_b-\phi_a\phi_b\right)D^b \, .
\end{equation}
This agrees with the memory effect formulas of Christodoulou \cite{Christodoulou} and Bieri and Garfinkle \cite{BG1}.

Let us now take the limit as $t_0 \to -\infty$ of the metric perturbation \eqref{hI}-\eqref{hIII}. It is clear from eqs.~\eqref{hI} and \eqref{hII} that 
\begin{equation}
\lim_{t_0 \to -\infty} (h^{\rm I}_0)_{ab} = \lim_{t_0 \to -\infty} (h^{\rm II}_0)_{ab} = 0 \, .
\end{equation}
On the other hand, $(h^{\rm III}_0)_{ab} = 4\varphi_0 k_a k_b$, so the limit as $t_0 \to -\infty$ of 
$(h^{\rm III}_0)_{ab}$ does not exist. We conclude that, as in the electromagnetic case, the retarded solution for the metric perturbation of a particle that moves forever on a null geodesic does not exist in the Lorenz gauge. On the other hand, the contribution of $(h^{\rm III}_0)_{ab}$ to the linearized Riemann tensor is
\begin{equation}
(R^{\rm III}_0)_{abcd} = 8\nabla_{[a} \nabla_{|[d} \varphi_0 k_{c]|} k_{b]}
\end{equation}
and it follows that the limit as $t_0 \to -\infty$ of $(R^{\rm III}_0)_{abcd}$ {\em does} exist. In fact, we obtain
\begin{equation}
R_{abcd} = \lim_{t_0 \to -\infty} (R^{\rm III}_0)_{abcd} = 4 k_{[a} \nabla_{b]} F_{cd}
\label{Rnull0}
\end{equation}
where $F_{ab}$ is given by \eqref{Fret} (with $q$ replaced by $E$) and the derivative is taken in the distributional sense. To calculate this distributional derivative more explicitly, let $\beta^{abcd}$ be smooth and of compact support and have the tensor symmetries of the linearized 
Riemann tensor. We wish to evaluate
\begin{equation}
-16\int \nabla_b \beta^{abcd} k_a \left(-\frac{1}{\rho}\rho_c k_d\right) \rho d\rho d\phi dz \, .
\end{equation}
To do so, we exclude a disc of radius $\epsilon$ about $\rho = 0$, integrate by parts with respect to $\rho$, and then let $\epsilon \to 0$. We thereby obtain 
\begin{equation}
R_{abcd}=16E\frac{1}{\rho^2}k_{[a}(\rho_{b]}\rho_{[c}-\phi_{b]}\phi_{[c})k_{d]}\delta(u) \, - \, 16\pi Ek_{[a}q_{b][c}k_{d]}\delta(x)\delta(y) \delta(u) \, ,
\label{Rnull}
\end{equation}
where $q_{ab}$ is the projection of the metric into the ``$x$-$y$'' plane.
Equation \eqref{Rnull} agrees with the Riemann curvature tensor of the Aichelburg-Sexl solution---apart from several sign discrepancies, which are undoubtedly misprints in eq.(3.12) of their paper\footnote{In particular, the Ricci component $R_{00}$ is easily computed by adding together the first two lines of their eq.(3.12) and does not agree with the (correct) expression they give below eq.(3.12); their eq.(3.12) also fails to be rotationally invariant in the plane orthogonal to the direction of the particle.}. Equation \eqref{Rnull} may be interpreted as the linearized curvature\footnote{As Aichelburg and Sexl have argued, this solution may be interpreted as a solution to the full, nonlinear Einstein equation, not merely the linearized Einstein equation. Indeed, Aichelberg and Sexl obtained their solution by taking an infinite boost limit of the exact Schwarszchild solution.} of the retarded field of particle of energy $E$ that moves on a null geodesic forever. Although, as we have seen, the retarded solution for the perturbed metric in the Lorenz gauge, does not exist as a distribution, it should be possible to find other gauges in which a distributional metric perturbation giving rise to \eqref{Rnull} does exist.

Unlike \eqref{RI+II+III}, the Riemann tensor \eqref{Rnull} does not have a derivative of a $\delta$-function term. Furthermore, its effects fall off at large distances like $1/r^2$ rather than $1/r$. Consequently, {\em we conclude there is no memory effect associated with the retarded field of a particle that moves on a null geodesic forever.} However, the $\delta$-function in \eqref{Rnull} will produce an instantaneous ``relative velocity kick'' to a system of test particles moving on geodesics. Integrating \eqref{Rnull}, we find that if the particles have initially separation $D^a$, the relative velocity kick will be 
\begin{equation}
\Delta v_a= 4E\frac{1}{\rho^2}\left(\rho_{a}\rho_{b} - \phi_{a}\phi_{b}\right)D^b \, .
\label{ASkick}
\end{equation}
This velocity kick can be given a simple interpretation in terms of Newtonian tidal effects. Consider, in Newtonian gravity, a particle of mass $E$ traveling with velocity $c$ along the $z$-axis. The Newtonian potential produced by such a particle at time $t$ is
\begin{equation}
\chi=-\frac{E}{\sqrt{x^2+y^2+(z-ct)^2}} \, .
\end{equation}
The tidal tensor associated with this potential is
\begin{equation}
\Phi_{a b}=\frac{E}{r'^3}(3r'_a r'_b-\delta_{ab}) \, ,
\end{equation}
where $r'=\sqrt{x^2+y^2+(z-ct)^2}$ and $r'_a = \nabla_a r'$. We can integrate the tidal tensor once to get the net relative velocity change of two neighboring test particles over all time. For test particles initially separated by the displacement $D^j$, we obtain
\begin{align}
\Delta v_i = \int_{-\infty}^\infty\textnormal{d}t'\Phi_{ij}(t',x,y,z)D^j\nonumber\\
=2E\frac{1}{\rho^2}(\rho_i \rho_j - \phi_i \phi_j)D^j \, .
\end{align}
Thus, apart from a factor of $2$, the net relative velocity change in the Newtonian case produced by a particle of mass $E$ that moves forever along the $z$-axis at velocity $c$ agrees with the relative velocity kick in linearized gravity produced by a particle of energy $E$ that moves forever on a corresponding null geodesic. The only difference is that in Newtonian gravity, these tidal effects occur over all time, whereas in linearized gravity, the tidal effects are ``compressed'' into a null plane traveling along with the source. Thus, the Newtonian tidal acceleration is gradual and continuous, whereas in linearized gravity, one obtains an instantaneous velocity kick.

\section{Summary and Conclusions}

We have have investigated the retarded solution for a scalar field, an electromagnetic field, and a linearized gravitational field associated with the creation of a null particle at time $t_0$ in Minkowski spacetime. In the scalar case, we can simply create a charged null particle; in the electromagnetic and linearized gravitational cases, other sources must also be present in order to conserve, respectively, charge and $4$-momentum. There are then two distinct limits of this retarded solution that we can take. The first is to fix $t_0$ and extract the leading order in $1/r$ behavior of the solution. In all three cases, there are effects produced on distant test particles at order $1/r$ caused by the creation of the null particle. In the scalar and electromagnetic cases, they give rise to an instantaneous ``kick'' to the $4$-momentum of a test particle. In the linearized gravitational case, the $O(1/r)$ effect is to produce an instantaneous relative displacement of test particles---the memory effect.

The alternative limit is to fix the observation point and let $t_0 \to -\infty$. This limit can be thought of as providing the retarded field of a null particle that moves on a null geodesic forever. In the scalar case, we found that this limit does not exist as a distribution. However, in the electromagnetic and linearized gravitational cases, although the limits of the Lorenz gauge vector potential and Lorenz gauge metric perturbation similarly do not exist, the limits of the electromagnetic field tensor and linearized Riemann tensor do exist. In the electromagnetic case, the limiting electromagnetic field tensor gives rise to a velocity kick on distant test particles at order $1/r$, but the form of this velocity kick is very different from the $O(1/r)$ velocity kick produced by the creation of a null charge at finite time $t_0$. In the linearized gravitational case, the limiting linearized Riemann tensor yields the Aichelberg-Sexl solution. It falls off as $1/r^2$ and thus produces no effects of any kind at order $1/r$. In particular, there is no memory effect. The leading order ($1/r^2$) effect of this linearized Riemann tensor is to produce an instantaneous relative velocity kick on test particles, of exactly the same form as the integrated Newtonian tidal force would produce.

We conclude that in linearized gravity, the ``radiation field'' (retarded solution) produced by a particle moving on a null geodesic forever is the Aichelberg-Sexl solution, which is a pure ``tidal field'' that produces no associated memory effect. Thus, the memory effect should not be interpreted as being caused merely by the passage of (effective) stress-energy to null infinity. However, as already noted in the Introduction, the Aichelberg-Sexl solution (even with a smoothed out source) fails to be asymptotically flat at spatial infinity, and thus is not physically acceptable. One way of producing a physically acceptable solution is to create the null particle at a finite time $t_0$ via an ``emission event,'' as we have considered. In that case, there will be a burst of radiation associated with the emission event that produces a nontrivial memory effect, in agreement with previous results. More generally, the requirement of asymptotic flatness at spatial infinity implies either the finite time creation of the null particle or the presence of additional ``incoming radiation'' from past null infinity that is not directly associated with the null particle. We believe that the memory effect is most naturally interpreted as being caused by either the emission event or by the additional incoming radiation from past null infinity, rather than by the passage of the particle to future null infinity.

\bigskip

\noindent
{\bf Acknowledgements}

We wish to thank Lydia Bieri and David Garfinkle for helpful discussions. This research was supported in part by NSF grant PHY12-02718 to the University of Chicago. Some of this research was carried out while one of us (R.M.W.) was in residence during September, 2013 at the Mathematical Sciences Research Institute in Berkeley, California, supported by the National Science
Foundation under Grant No. 0932078 000.

\end{document}